\documentclass[twocolumn, secnumarabic, amsmath, amssymb, nobibnotes, showkeys, aps, prl]{revtex4-2}

\usepackage{graphicx}
\usepackage{bm}
\usepackage{caption}
\usepackage{subcaption}
\usepackage{natbib}
\usepackage{tikz}
\usepackage{hyperref}
\usepackage[normalem]{ulem}
\usepackage{overpic}
\usepackage{lipsum}

\hypersetup{
    colorlinks=true,
    linkcolor=blue,
    filecolor=blue,      
    urlcolor=blue,
    citecolor=blue
    }

\def\uu{\mathbf{u}}

\begin{document}

\title{
Bistable flow dynamics of airfoil stall under varying angle of attack:\\ 
A stochastic model with multiplicative noise
}

\author{Edouard Boujo} 
\affiliation{Laboratory of Fluid Mechanics and Instabilities, École Polytechnique Fédérale de Lausanne, CH-1015 Lausanne, Switzerland} 

\author{Ivan Kharsansky Atallah}
\affiliation{Fluid Mechanics Department, ENSTA Paris, Institut Polytechnique de Paris, F-91120 Palaiseau, France} 
\affiliation{EM2C Laboratory, CNRS, CentraleSupelec, Universit\'e Paris-Saclay, F-91190 Gif-sur-Yvette, France} 

\author{Luc R. Pastur}
\affiliation{Fluid Mechanics Department, ENSTA Paris, Institut Polytechnique de Paris, F-91120 Palaiseau, France} 


\begin{abstract}

We focus on the intermittent bistable stall dynamics of an airfoil under varying angle of attack. We propose a one-dimensional Langevin equation where the stochastic forcing depends on the state of the system -- high-lift attached flow or low-lift detached flow -- and where the deterministic potential depends continuously on the angle of attack. The model, identified based on the flow statistics and dynamics, reproduces the S-shaped lift curve, as well as the flow dynamics. It also predicts the nature of the bifurcations that the flow undergoes as the angle of attack varies. 

\end{abstract}


\maketitle


Systems that exhibit multistable dynamics  with transitions between two or more states are ubiquitous:
biological phenomena \cite{goldbeter2018} such as cell division, differentiation, cancer onset, and apoptosis \cite{wilhelm2009},  ecological systems \cite{guttal2007},
and many fluid dynamical systems such as transition to turbulence \cite{barkley2015nature}, 
turbulent liquid metal flows \cite{berhanu2007prl}, 
reversal of the Earth's magnetic field \cite{jacobs1994book, Petrelis2009},  
turbulent Rayleigh-B\'enard convection \cite{Araujo2005,Brown2007},
forced turbulent shear flows \cite{Dallas2020},
climate change \cite{bathiany2018},  
thermoacoustic instabilities \cite{Noiray_Schuermans_2013, Gopalakrishnan_Sujith_2015},
wake flows behind spheres \cite{grandemange2014exif} or  more complex objects \cite{Grandemange12PRE, rigas2015jfm, mallat2021jweia, Gayout2021} at large Reynolds numbers, or, as recently observed, intermittent dynamics of wing stall \cite{kharsansky2024}. 
While identifying simple reduced-order models able to capture the observed dynamics is crucial to understand, predict and control natural and technological phenomena, the task is challenging for
multistable dynamics with random, noise-induced transitions. 
Such systems are usually understood and modeled as being both (i)~strongly structured around attracting coherent states, giving it its deterministic character, and (ii)~subject to stochastic forcing due to a strongly unsteady environment  evolving on  timescales much shorter than those of the deterministic part of the dynamics.
The Langevin equation, a stochastic differential equation, is a celebrated class of such models  \cite{FRIEDRICH201187, Anvari2016, kwok2018book}.
In most fluid dynamical configurations, the noise is assumed to be additive, i.e. independent of the flow state. 
This is usually justified, especially when the structuring states are related through a broken symmetry of the system, as for instance in wake flows behind symmetric bluff bodies \cite{grandemange_gohlke_cadot_2013, parezanovic2015, rigas2015jfm}. 
However, the use of state-dependent multiplicative noise may improve the model, as observed in \cite{callaham2021prsa}. 
This is especially true when the  structuring states of the system differ substantially, like  in wing stall dynamics: the high-lift flow is essentially attached to the wing, while the low-lift flow is massively detached \cite{kharsansky2024}.
In addition, most studies  usually focus on a single operating point of the system, for example at a fixed Reynolds number or angle of incidence, whereas the dynamics unfold when the control parameters vary, which can lead to bifurcation points \cite{brunton_discovering_2016}.

In this Letter, we propose a method for describing the evolution of  bistable dynamics in the presence of a varying control parameter, while changing the noise intensity according to the state of the system. 
We apply the method to the flow past a stalled airfoil, which exhibits, over a finite range of  angle of incidence, intermittent dynamics between  states of high and low lift. 
The low-lift  state exhibits flow fluctuations of greater amplitude than the high-lift state, which suggests introducing a stochastic forcing dependent on the lift state. 
In addition, our model  accounts for variations in angle of incidence, including the  bistable zone of the stall dynamics. 
Based on the model, we draw conclusions about the nature of the bifurcation points delimiting this bistable zone.

\begin{figure}[t] 
\centerline{
  \begin{overpic}[width=0.45\textwidth, trim=0mm 0mm 0mm 0mm, clip=true]{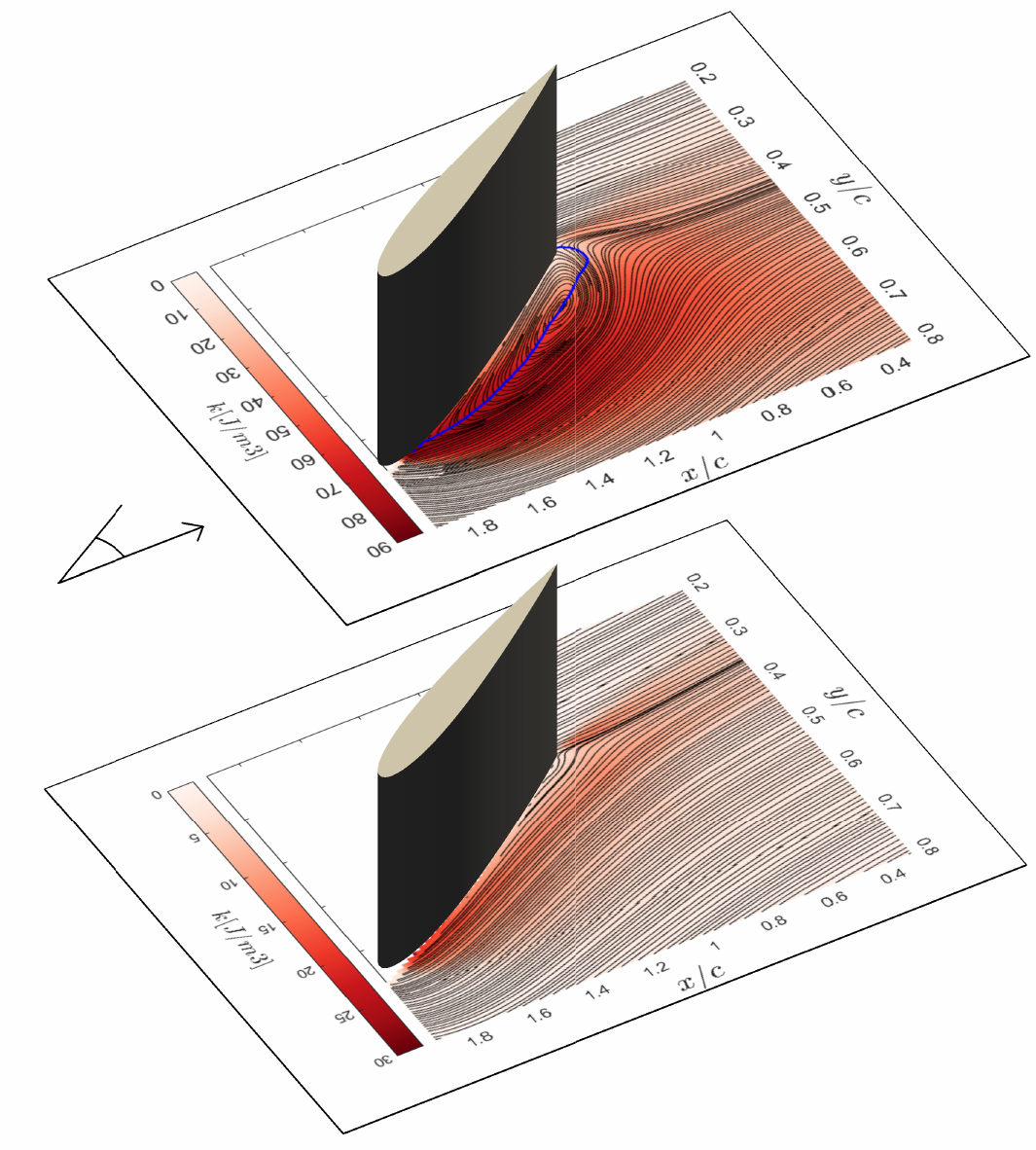}
    \put(5,85){\textcolor{red}{State D}}
    \put(8,46){$U_\infty$}
    \put(10,53){$\alpha$}
    \put(7,9.5){\textcolor{blue}{State A}}
  \end{overpic}
}
\caption{States A and D from conditional particle image velocimetry for an angle of attack $\alpha$ in the bistable range
and $Re=1.51\times 10^5$.
In state A, the boundary layer is attached to the wing, while the flow is detached in state D, as revealed by the region of reverse flow (blue line) 
and the near-wall recirculation bubble. 
Black: streamlines;
contours: turbulent kinetic energy $k$ (in J/m$^3$).
Axes expressed in units of wing chord $c$.
}
\label{fig:naca} 
\end{figure}

\begin{figure}
\centerline{
   \begin{overpic}[height=7.6cm, trim=59mm 0mm 0mm 0mm, clip=true]{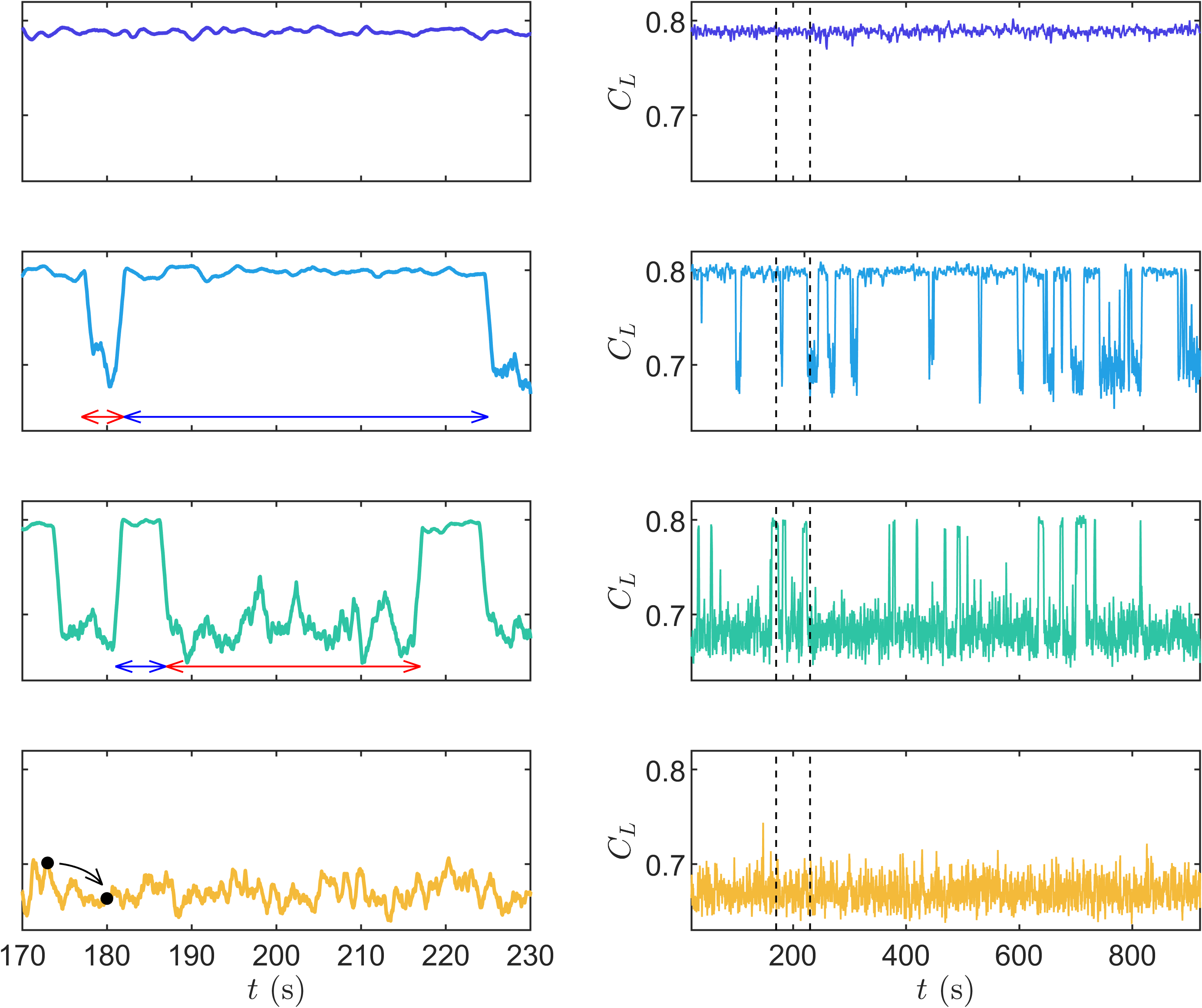}
      \put(-2,98){$(a)$}
   \end{overpic}    
   \hspace{0.1cm}
   \begin{overpic}[height=7.6cm, trim=0mm 0mm 63mm 0mm, clip=true]{figures/exp_signals-v2.png}
     \put(-3.5,98){$(b)$}
     \put(4,61.2){\textcolor{red}{\scriptsize $t_{res}^{\mathrm{D}}$}}
     \put(27,61.){\textcolor{blue}{\scriptsize $t_{res}^{\mathrm{A}}$}}
     \put(20,93){\textcolor[rgb]{0.2775    0.2499    0.8920}{\scriptsize $\alpha=10.70^\circ$}}
     \put(20,69){\textcolor[rgb]{0.1390    0.6259    0.8981}{\scriptsize $\alpha=10.82^\circ$}}
     \put(20,46){\textcolor[rgb]{0.1804    0.7701    0.6447}{\scriptsize $\alpha=10.90^\circ$}}
     \put(20,21){\textcolor[rgb]{0.9576    0.7285    0.2285}{\scriptsize $\alpha=11.02^\circ$}}
     \put(4.5,15.7){\scriptsize $t$}
     \put(10,14){\scriptsize $t+\tau$}
  \end{overpic}  
}  
\caption{
Transitions between attached (A) and detached (D) flow states:
lift coefficient $C_L$ measured experimentally for different angles of attack $\alpha$.
$(a)$~Fifteen minutes of the one-hour long signals. 
$(b)$~Zoom on the one-minute interval shown with dashed lines in $(a)$.
Horizontal blue and red arrows show a few residence times $t_{res}$ in states A and D. 
Jump probabilities over a time shift $\tau$ (black arrow) are used to compute finite-time Kramers-Moyal coefficients  [Eq.~(\ref{eq:KM})].
}
\label{fig:stat} 
\end{figure}
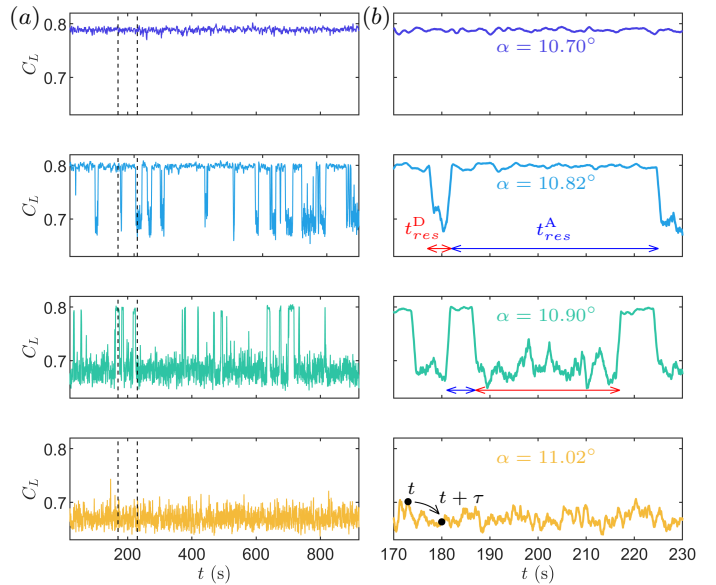

\textit{Experimental configuration} -- 
We perform experiments on a  thin symmetric NACA0012 airfoil (Fig.~\ref{fig:naca}) at a Reynolds number $Re=\rho cU_\infty/\mu \simeq 1.16\times10^5$, based on the wing chord $c=120$ mm, freestream velocity $U_\infty$,  air density $\rho$ and dynamic viscosity $\mu$.
The turbulence intensity in the wind tunnel is 0.4\%.
We use the lift coefficient $C_L = 2L/(\rho U_\infty^2cs)$ as a global scalar observable of the state of the system,  
where $L$ is the lift force and $s=450$ mm the wing span. 
One-hour long signals $C_L(t)$ were obtained with a rotating balance measuring $L$ (resolution $\pm 1$ mN at 1 kHz) for several angles of attack   $\alpha\in[10.66^\circ, 11.10^\circ]$ (resolution $\pm 0.02^\circ$) \cite{kharsansky2024}. 
Special care was taken in the choice of the acquisition time as the model identification method relies on converged statistics.
Times series of $C_L(t)$ (Fig.~\ref{fig:stat}) exhibit intermittent dynamics between two states of high and low lift, with random and abrupt transitions. 

Figure~\ref{fig:naca} shows the two mean-flow states, the high-lift attached (A) state and the low-lift detached (D) state, 
obtained via $C_L$-based conditional averaging of  instantaneous velocity fields produced by particle image velocimetry (double pulse Nd:YAG laser Litron Nano T 135, output energy 135~mJ at 532~nm, maximum repetition rate 15~Hz, beam diameter 5~mm) \cite{thesisKharsansky}.
In the D state, a recirculation bubble typical of  detached flows is  visible on the wing suction side (blue line), while in the A state the  flow remains mostly attached.

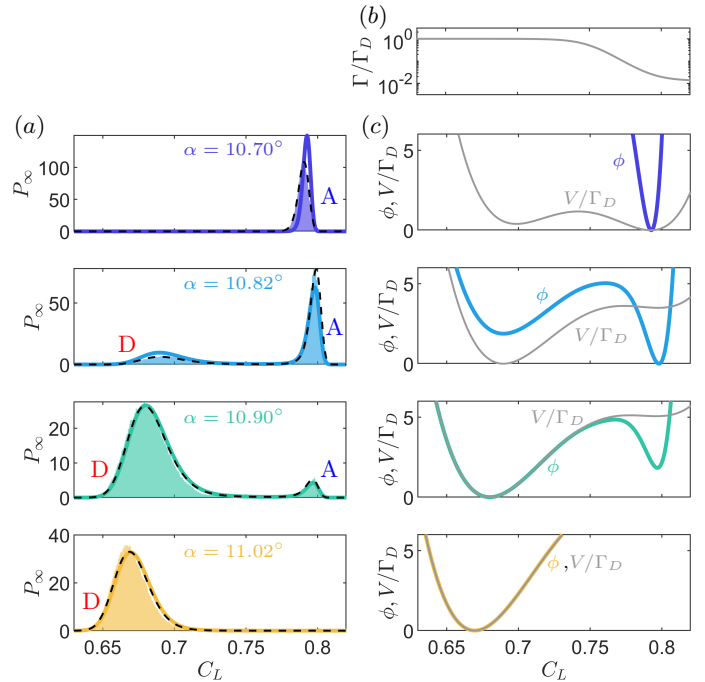
\begin{figure}
\centerline{
  \hspace{4.4cm}
  \begin{overpic}[width=4.5cm, trim=0mm 8mm 0mm 0mm, clip=true]{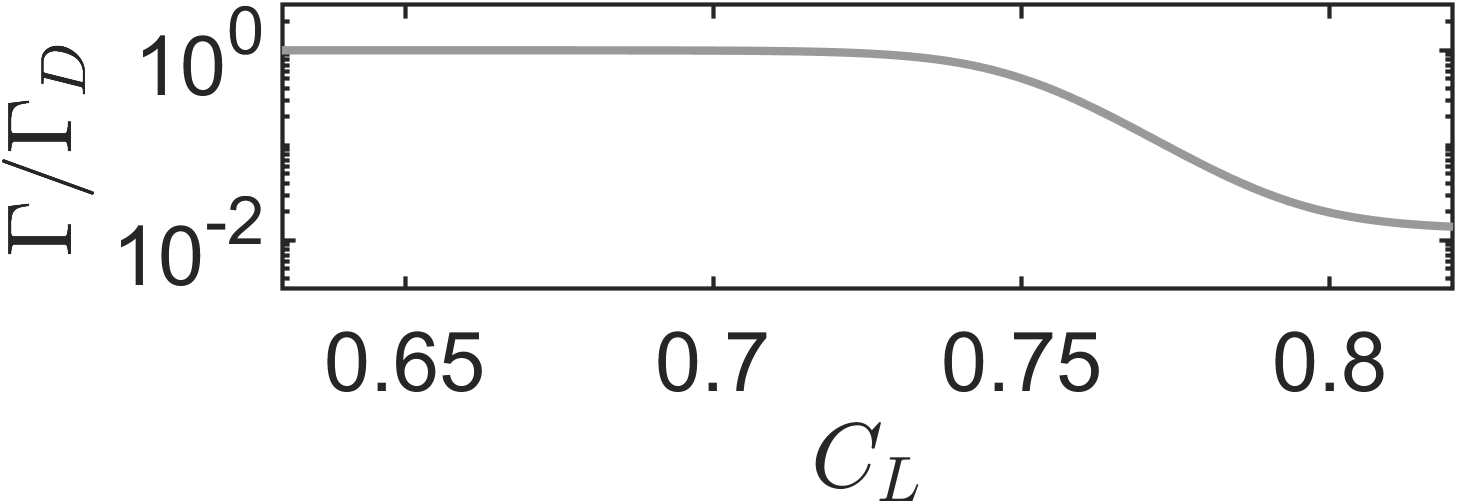}
     \put(2,23){$(b)$}
  \end{overpic} 
}  
\vspace{0.4cm}
\centerline{
  \begin{overpic}[height=7.273cm, trim=0mm 0mm 0mm 0mm, clip=true]{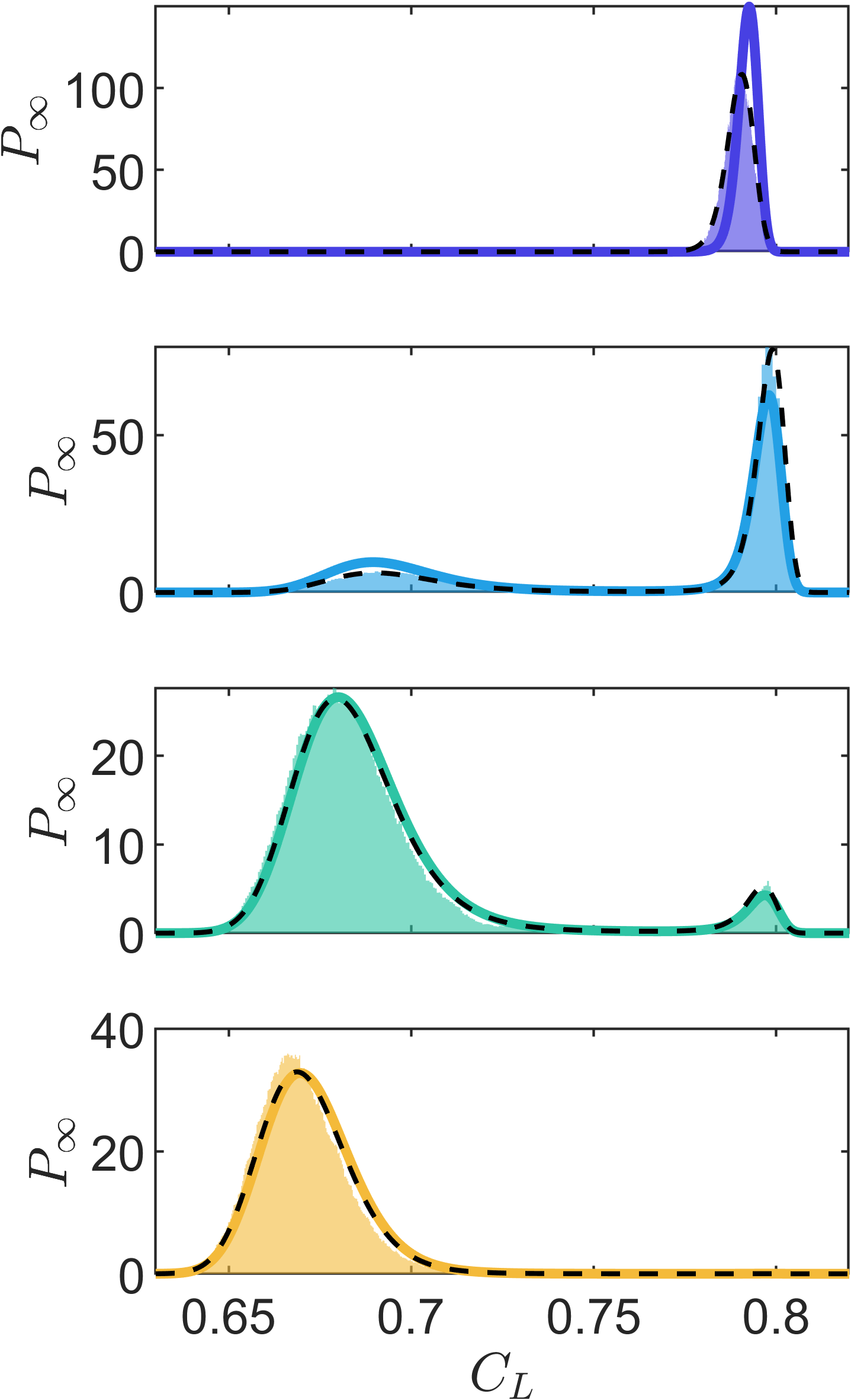}
     \put(0,100 ){$(a)$}
     \put(31,96  ){\textcolor[rgb]{0.2775    0.2499    0.8920}{\scriptsize$\alpha=10.70^\circ$}}
     \put(31,71.5){\textcolor[rgb]{0.1390    0.6259    0.8981}{\scriptsize$\alpha=10.82^\circ$}}
     \put(31,47  ){\textcolor[rgb]{0.1804    0.7701    0.6447}{\scriptsize$\alpha=10.90^\circ$}}
     \put(31,22.5){\textcolor[rgb]{0.9576    0.7285    0.2285}{\scriptsize$\alpha=11.02^\circ$}}
     \put(19,  60){\textcolor{red}{D}}
     \put(14,  37){\textcolor{red}{D}}
     \put(12.5,13){\textcolor{red}{D}}
     \put(56,87){\textcolor{blue}{A}}
     \put(57,63){\textcolor{blue}{A}}
     \put(56,37){\textcolor{blue}{A}}
  \end{overpic} 
  \hspace{0.25cm} 
  \begin{overpic}[height=7.273cm, trim=0mm 0mm 0mm 0mm, clip=true]{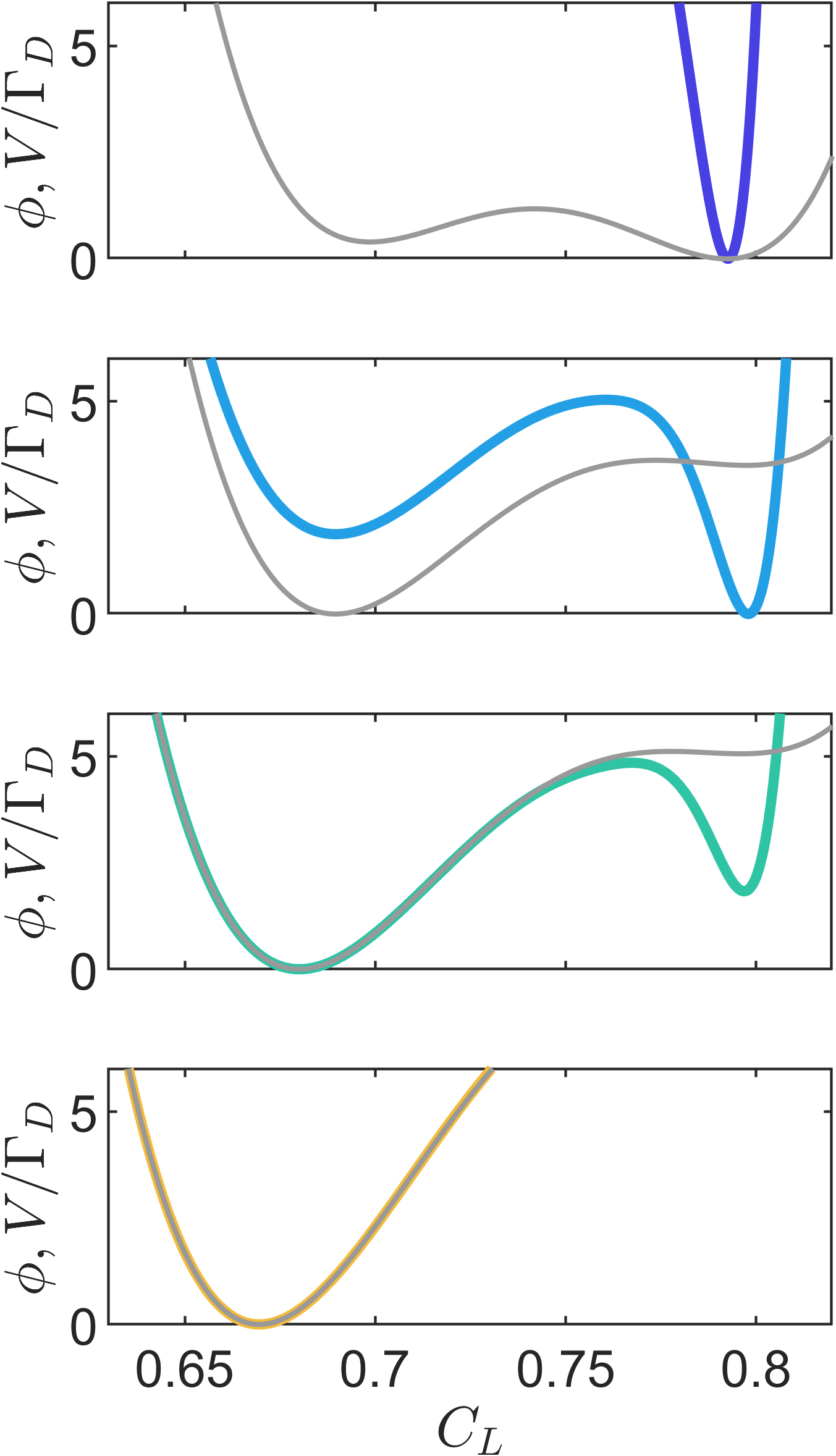}
     \put(-3,100 ){$(c)$}
     \put(43,94  ){\textcolor[rgb]{0.2775    0.2499    0.8920}{\scriptsize$\phi$}}
     \put(29,69.5){\textcolor[rgb]{0.1390    0.6259    0.8981}{\scriptsize$\phi$}}
     \put(31,38  ){\textcolor[rgb]{0.1804    0.7701    0.6447}{\scriptsize$\phi$}}
     \put(31,20.5){\textcolor[rgb]{0.9576    0.7285    0.2285}{\scriptsize$\phi$}}
     \put(34,87  ){\textcolor[rgb]{0.6 0.6 0.6}{\scriptsize$V/\Gamma_D$}}
     \put(36,62  ){\textcolor[rgb]{0.6 0.6 0.6}{\scriptsize$V/\Gamma_D$}}
     \put(28,47  ){\textcolor[rgb]{0.6 0.6 0.6}{\scriptsize$V/\Gamma_D$}}
     \put(34,20.5){\scriptsize,\textcolor[rgb]{0.6 0.6 0.6}{\scriptsize$V/\Gamma_D$}}
  \end{overpic}
}
\caption{
$(a)$~Lift PDF. 
Shaded histogram: experimental $P_\infty(C_L)$.
Dashed line: individual fit $P_\infty(C_L)$.
Solid line: global model $P_\infty(C_L;\alpha)$.
$(b)$~Identified noise intensity $\Gamma$.
$(c)$~Identified potential-to-noise ratios $V/\Gamma_D$ (thin gray line) and $\phi$ (thick colored line). 
Same color code as in Fig.~\ref{fig:stat}.
}
\label{fig:pdf} 
\end{figure}

At smaller angles of attack, state A is more probable and has a longer lifetime than state D [Fig.~\ref{fig:stat}(a)-(b)], while state D becomes more probable with longer lifetimes at larger angles of attack [Fig.~\ref{fig:stat}(c)-(d)]. 
The two states are equally probable at a critical value $\alpha_c \approx 10.85^\circ $. 
This transition of the more probable state from A to D as $\alpha$ increases
is also apparent in the probability distribution functions (PDFs) of Fig.~\ref{fig:pdf}. 
We note in these PDFs that the peak of state A is narrower than that of state D, which means that lift fluctuations are smaller in state A than in state D. For modeling purposes, it will therefore be necessary to change the stochastic forcing in each of the two states. 

\textit{Model} --
In our wing stall experiments, 
$C_L(t)$ is a stochastic Markov process
with memoryless dynamics, random switches and residence times  larger than the convective time $c/U_\infty$ by orders of magnitudes 
(see Supplemental Material [\textit{URL will be inserted by publisher}] for evidence of the Markov property).
We  describe the behavior of the flow, which consists of infinitely many degrees of freedom, via a low-dimensional model, specifically
a one-dimensional Langevin equation for $C_L(t)$,
\begin{align}
\dot C_L= F(C_L) + \sqrt{2\Gamma(C_L)} \xi(t).
\label{eq:SDE_CL}
\end{align}
The drift coefficient $F(C_L) =  -\mathrm{d} V/\mathrm{d}C_L$ represents the deterministic part and derives from a potential $V$.
The diffusion coefficient $\Gamma(C_L)$ and the Gaussian white noise term $\xi(t)$ (characterized by $\langle\xi(t)\rangle=0$ and $\langle \xi(t) \xi(t') \rangle = \delta(t-t')$) represent the stochastic part. 
We choose  a simple quartic  potential able to capture bistability,
\begin{align}
V(C_L) =  a C_L + \frac{b}{2} C_L^2 + \frac{c}{3} C_L^3 + \frac{d}{4} C_L^4,
\label{eq:V}
\end{align}
where the unknown coefficients $a,b,c,d$ depend on the angle of attack $\alpha$. 
Anticipating on the results, we find that additive noise ($\Gamma$ independent of $C_L$) is  not able to reproduce the observed statistics 
(see also Supplemental Material [\textit{URL will be inserted by publisher}]), whereas multiplicative noise $\Gamma(C_L)$ yields excellent results. 
Because we expect the  noise intensity to differ substantially in the attached and detached states, we use 
a  piecewise-constant noise intensity equal to $\Gamma_{A}$ and $\Gamma_{D}$ in states A and D, respectively, and varying smoothly in between,
\begin{align}
\Gamma(C_L) = \Gamma_D + \frac{\Gamma_A-\Gamma_D}{2} \left[ 1 + \tanh\left( \frac{C_L - C_L^*}{\Delta} \right) \right], 
\label{eq:G}
\end{align}
where $C_L^*=0.75$ corresponds to the PDF minimum -- the boundary between states A and D -- at all angles of attack 
and $\Delta=0.02$ is a small characteristic width [Fig.~\ref{fig:pdf}$(b)$].
The dependence of $\Gamma$ on $\alpha$ is expected to be weak in the narrow range of angles of attack investigated and is therefore neglected.
Hereafter, we use Ito’s interpretation of stochastic integrals \cite{risken_fokkerplanck_1984}. 

The evolution of the PDF $P(C_L,t)$ is governed by the Fokker-Planck equation
\begin{align}
\partial_t P = -\partial_{{C_L}} \left( F P \right)
+ \partial_{{C_L C_L}} \left( \Gamma P \right), 
\label{eq:FPE}
\end{align}
and the stationary PDF $\lim_{t\to \infty} P({C_L},t)$ is
\begin{align}
P_\infty({C_L}) 
=
\frac{\mathcal{N}}{\Gamma({C_L})} \exp\left( \int^{{C_L}} \frac{F(C_L')}{\Gamma(C_L')} \mathrm{d}C_L'  \right), 
\label{eq:stat_PDF}
\end{align}
with $\mathcal{N}$ a normalization factor such that $\int_0^\infty P_\infty({C_L}) \, \mathrm{d}{C_L} = 1$.
It is convenient to rewrite 
\begin{align}
P_\infty({C_L}) 
= 
\mathcal{N} \exp\left( -\phi(C_L) \right),
\label{eq:stat_PDF_phi} 
\end{align}
and to compare this expression with the stationary PDF for additive noise (constant intensity $\Gamma$),
\begin{align}
P^a_\infty({C_L})
=
\frac{\mathcal{N}}{\Gamma} \exp\left(  \frac{\int^{{C_L}} F(C_L') \mathrm{d}C_L'}{\Gamma} \right)
=
\mathcal{N}' \exp\left( - \frac{V}{\Gamma}  \right),
\label{eq:stat_PDF_add}
\end{align}
which shows that
\begin{align}
\phi(C_L) = \ln \left( \Gamma(C_L) \right) - \int^{{C_L}} \frac{F(C_L')}{\Gamma(C_L')} \mathrm{d}C_L'
\end{align}
is the equivalent of the standard potential-to-noise ratio $V/\Gamma$ in the case of constant noise intensity.


\begin{figure*}
\centerline{
  \begin{overpic}[height=4.5cm, trim=0mm 0mm 16mm 0mm, clip=true]{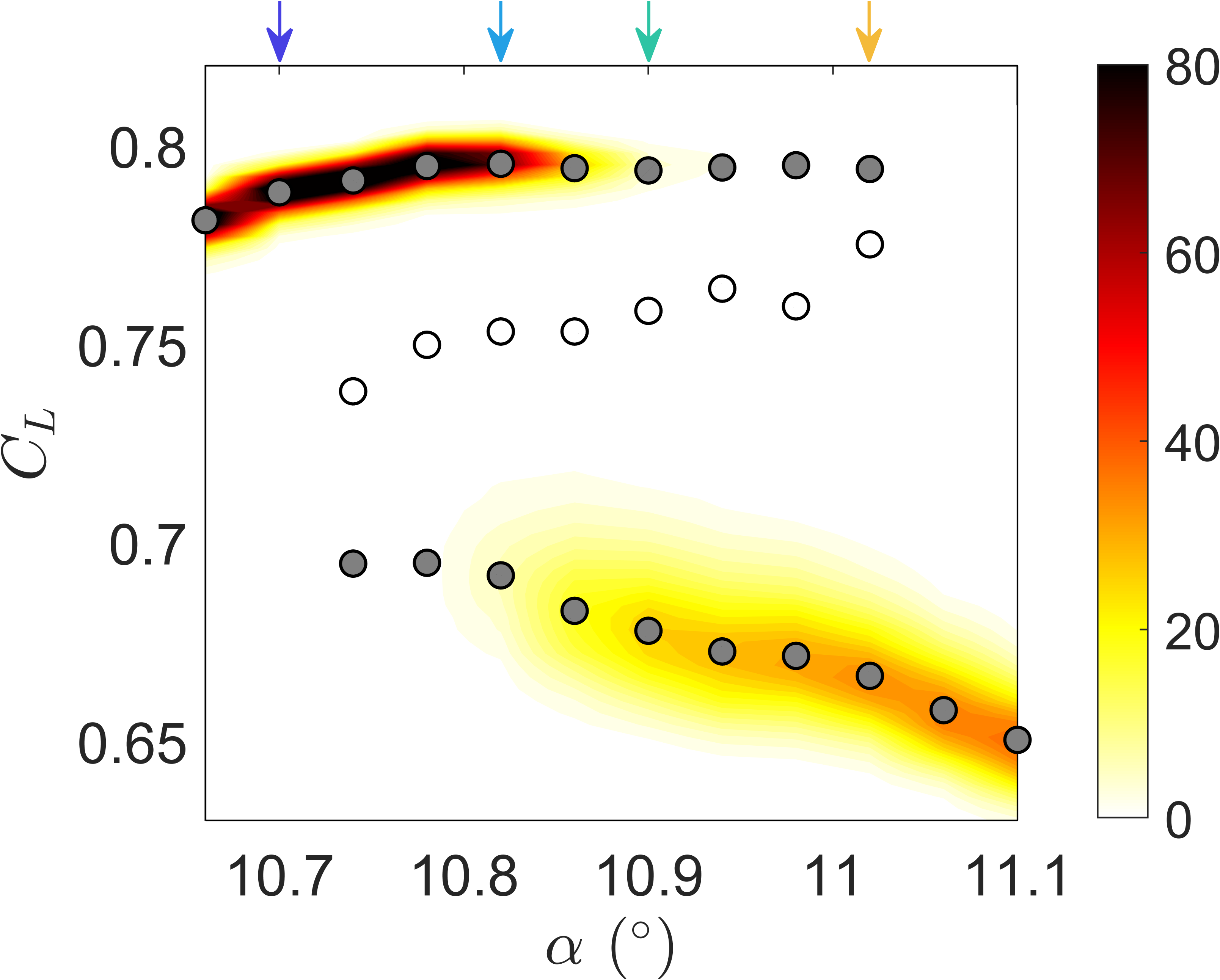}
    \put(-1,90.5){$(a)$}
    \put(68,78){\textcolor{blue}{A}}
    \put(35,30){\textcolor{red}{D}}
    \put(22,17.5){Experiment}
  \end{overpic} 
  \begin{overpic}[height=4.5cm, trim=21mm 0mm 0mm 0mm, clip=true]{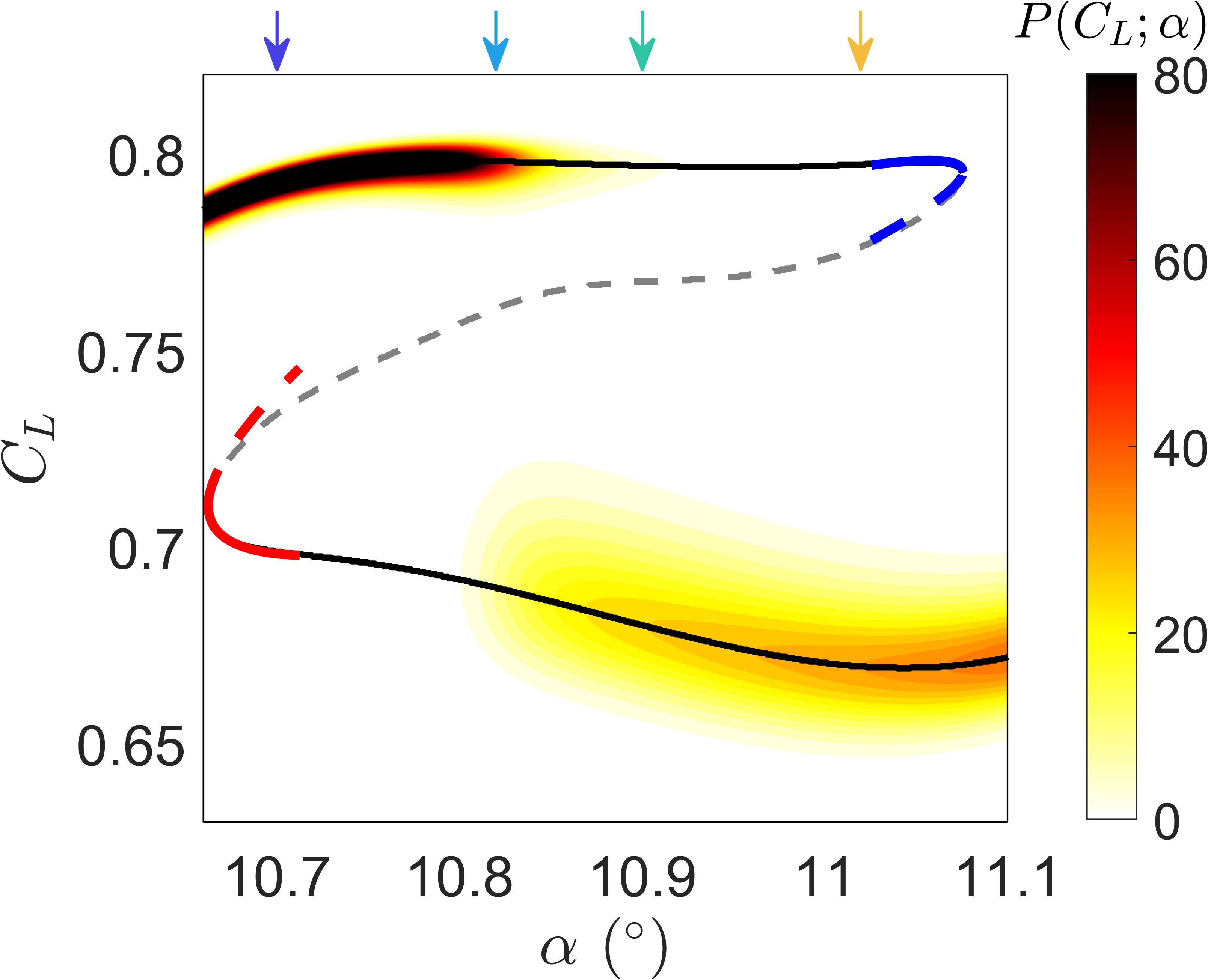}
    \put(-5,96){$(b)$}
    \put(52,81){\textcolor{blue}{A}}
    \put(21,32){\textcolor{red}{D}}
    \put(7,18){Model}
  \end{overpic}  
  \hspace{0.2cm}
  \begin{overpic}[height=4.5cm, trim=0mm 0mm 0mm 0mm, clip=true]{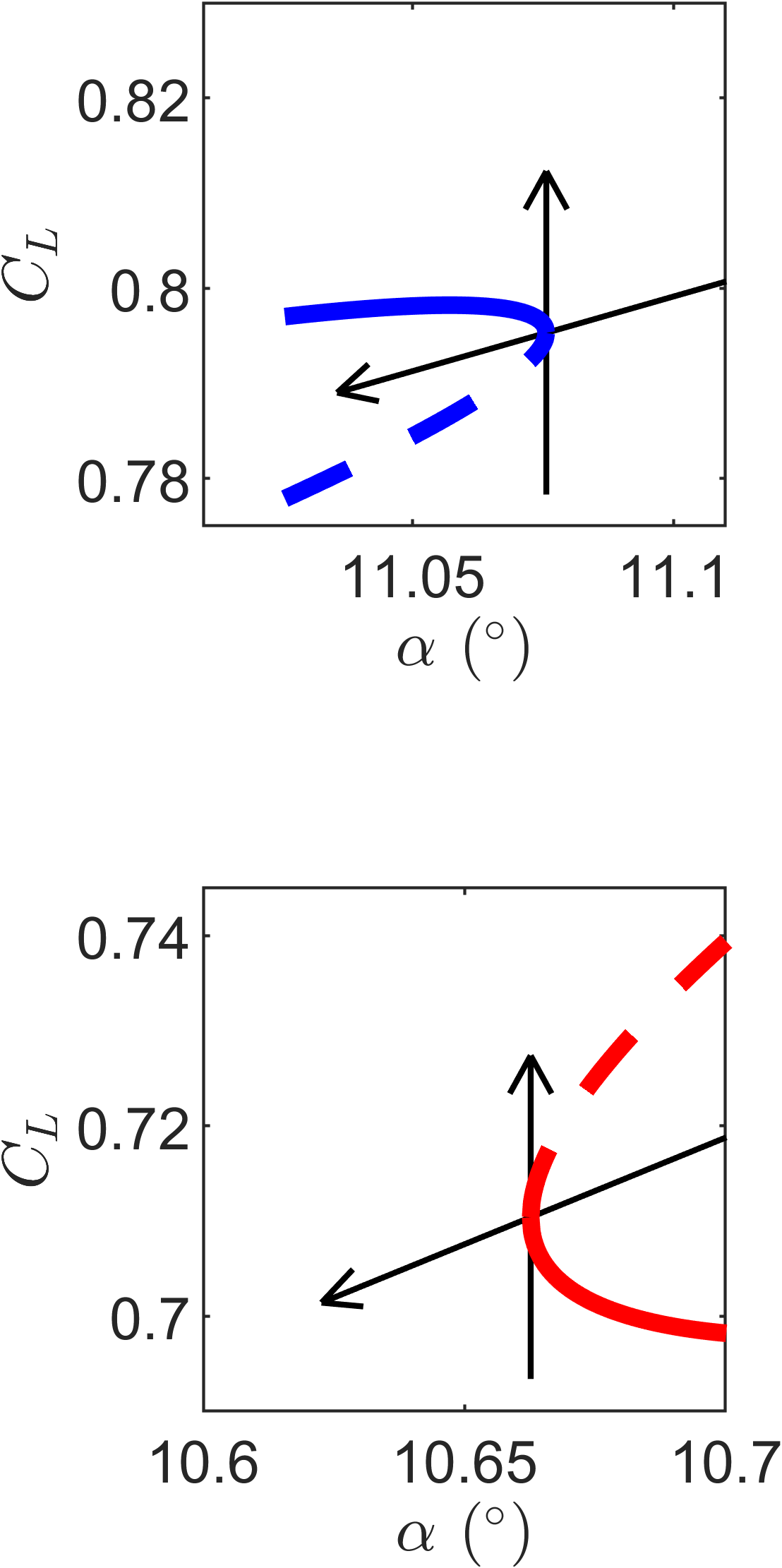}
     \put(-2,99.8){$(c)$}
     \put(14,  96){\scriptsize $s=-1$}
     \put(33.5,92){\scriptsize $\eta$}
     \put(17,  73){\scriptsize $\beta$}
     \put(14,  38.5){\scriptsize $s=1$}
     \put(32.5,34){\scriptsize $\eta$}
     \put(16,  14){\scriptsize $\beta$}
  \end{overpic} 
  \begin{overpic}[height=4.5cm, trim=0mm 0mm 0mm 0mm, clip=true]{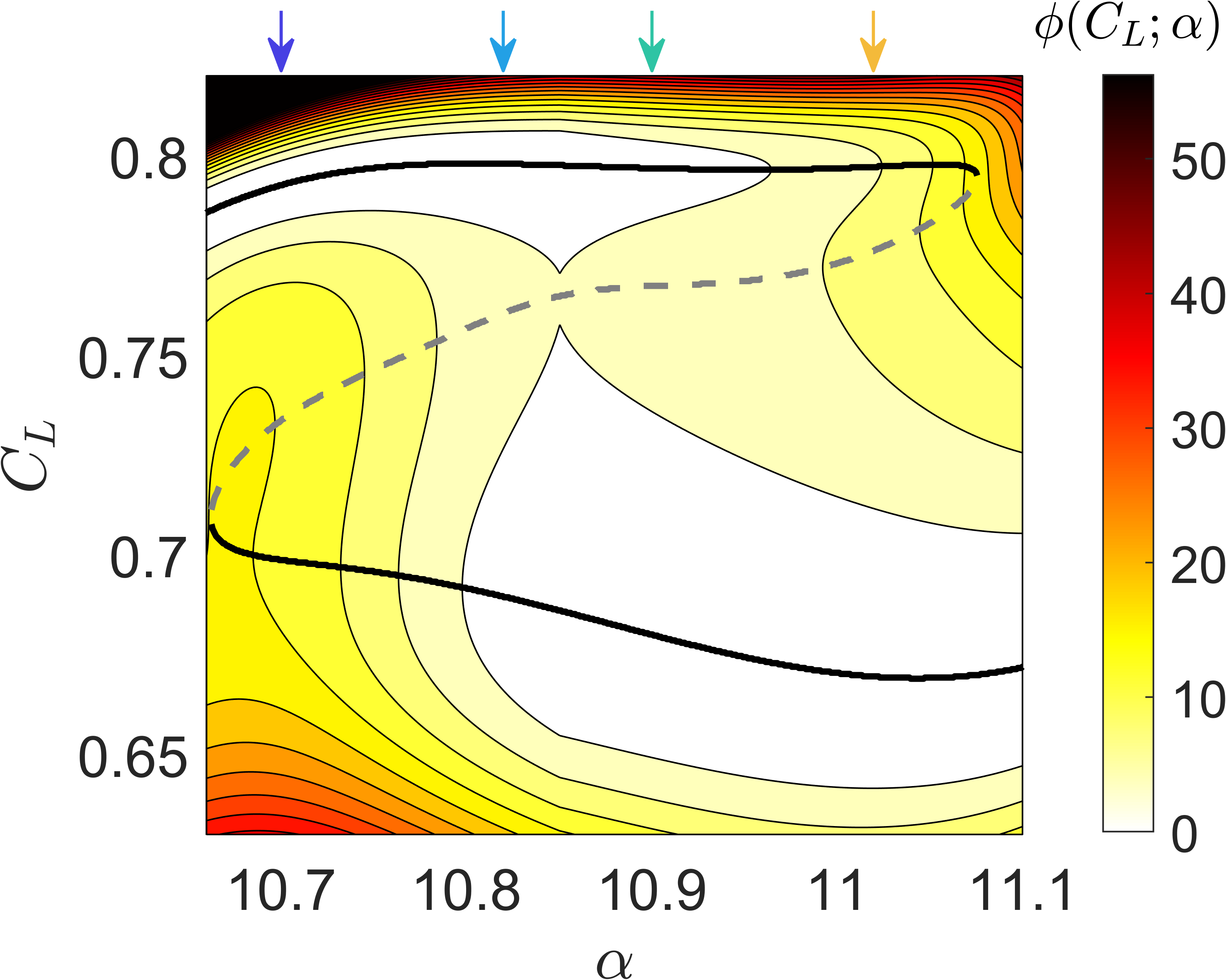}
    \put(0,80){$(d)$}
  \end{overpic}
  \hspace{0.1cm}
}
\caption{
$(a)$-$(b)$~Experimental and identified  PDF of the lift coefficient $C_L$ for varying  angle of attack $\alpha$ close to the bistable region.
Filled/open circles:   maxima/minima of $P_\infty(C_L;\alpha)$. 
Blue/red lines:  local saddle-node bifurcations described by the normal form expansion \eqref{eq:nf} around $\alpha _A=11.07^\circ$ and $\alpha _D=10.66^\circ$, respectively. 
Solid/dashed lines: stable/unstable deterministic branches. 
The colored arrows on the top axis mark the values of $\alpha$ used in Figs.~\ref{fig:stat}-\ref{fig:pdf}, 
with the same color code.
$(c)$~Zoom on the local saddle-node bifurcations.
See the main text and Eq.~\eqref{eq:nf} for the definition of $\beta$, $\eta$ and $s$.
$(d)$~Identified potential-to-noise ratio $\phi$, from which $P_\infty(C_L;\alpha)$ can be deduced directly with Eq.~\eqref{eq:stat_PDF_phi}.
}
\label{fig:s-shape}
\end{figure*}
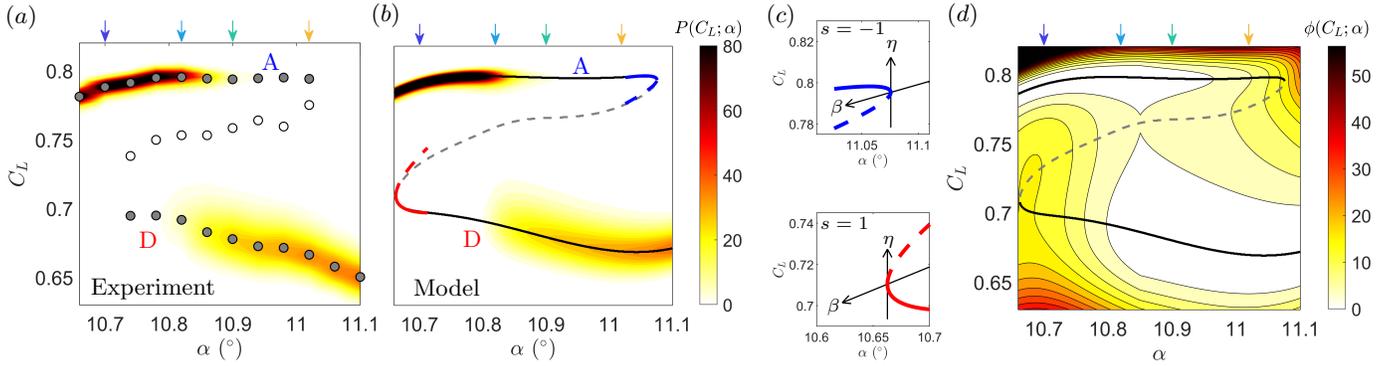

The model (\ref{eq:SDE_CL})-(\ref{eq:G}) consists of unknown $\alpha$-dependent coefficients $\{a,b,c,d\}$ and constant parameters $\{\Gamma_A,\Gamma_D\}$, which are now identified in two successive steps, first from the statistics of $C_L(t)$ and then from its dynamics. 

\textit{Statistics} --
Noting that $P_\infty({C_L})$ does not change if  the potential and the noise intensity are rescaled by the same factor, we identify the ratios $\{a,b,c,d,\Gamma_A\}/\Gamma_D$  for each $\alpha$ by finding the best fit of (\ref{eq:stat_PDF}) to the experimental PDF.
As shown in Fig.~\ref{fig:pdf}(a) for four  values of $\alpha$, the identified PDFs (black dashed lines)   are in excellent agreement with the experimental PDFs (shaded histograms).
Notably, the multiplicative-noise model  simultaneously captures the large fluctuations of state D and the smaller fluctuations of state A;
as shown in the Supplemental Material [\textit{URL will be inserted by publisher}], the additive-noise model cannot.
Also, Eq.~(\ref{eq:stat_PDF_add}) shows that, if the noise was additive,  the stationary PDF could be inferred directly from the  deterministic potential-to-noise ratio $V/\Gamma$ (gray lines in Fig.~\ref{fig:pdf}(c) for e.g. $\Gamma=\Gamma_D$), with minima of $V/\Gamma$ corresponding to maxima of $P_\infty$;
here, due to the multiplicative nature of the noise $\Gamma(C_L)$ [Fig.~\ref{fig:pdf}(b)], $P_\infty$ is instead in direct correspondence with the equivalent potential-to-noise ratio $\phi$ (colored lines in  Fig.~\ref{fig:pdf}(c)).

\textit{Dynamics} --
At this stage, the model coefficients $\{a,b,c,d,\Gamma_A\}$ are identified up to a multiplicative factor $\Gamma_D$.
To build a uniquely defined model, we  determine $\Gamma_D$ for each $\alpha$ from the dynamics of the system, i.e. from the time signals $C_L(t)$.
Considering the conditional probability $P(C_L',t' | C_L, t)$  that the signal is $C_L'$ at time $t'=t+\tau$ given that it was $C_L$ at time $t$ (arrow sketch in Fig.~\ref{fig:stat}), we compute finite-time Kramers-Moyal coefficients
\begin{align}
D^{(n)}_\tau(C_L) = \dfrac{1}{n!\tau} \int_{-\infty}^\infty (C_L'-C_L)^n P(C_L',t' | C_L, t) \, \mathrm{d}C_L'
\label{eq:KM}
\end{align}
for $n=1,2$ \cite{PhysRevLett.78.863, FRIEDRICH201187}.
For each $\alpha$, we obtain $\Gamma_D$ by minimizing the error between the experimental $D^{(n)}_\tau$ and those predicted by the adjoint Fokker-Planck equation for multiple time shifts $\tau$ and amplitudes $C_L$ \cite{honisch_estimation_2011, boujo_robust_2017}. 
We then  retain the mean value {as a single, $\alpha$-independent $\Gamma_D$.}

\begin{figure}
\vspace{0.2cm}
\centerline{  
  \begin{overpic}[height=5.5cm, trim=2mm 0mm 0mm 8mm, clip=true]{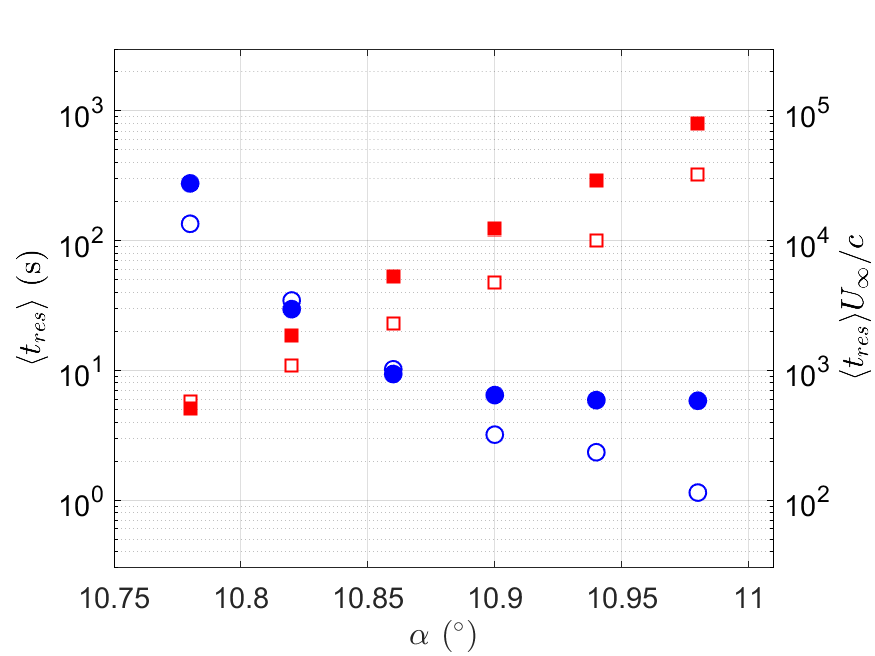}
    \put(20, 58){\textcolor{blue}{State A}}
    \put(54, 58){\textcolor{red}{State D}}
  \end{overpic}  
}
\caption{
Mean residence times of states A (circles) and D (squares) in physical/convective time units (left/right-hand axes). Open symbols: experiment; filled symbols: model. 
}
\label{fig:residence_time}
\end{figure}

\textit{Global parameterized model} --
Finally, continuous expressions of $\{a,b,c,d\}$ valid for all angle of attacks are constructed with  fits through values identified for each individual $\alpha$,
\begin{equation}
\log_{10} a(\alpha) =  a_0 + a_1 \alpha  + a_2 \alpha^2 + a_3 \alpha^3,
\label{eq:coeffs}
\end{equation}
and similarly for 
$b(\alpha)$, $c(\alpha)$, $d(\alpha)$.
The resulting global PDF $P_\infty(C_L;\alpha)$ is compared with the experimental one in Fig.~\ref{fig:s-shape}$(a)$-$(b)$. 
The model correctly reproduces the global shape (S-curve), the bistability region, and the distinct fluctuation amplitudes in states A and D. 
The model can also be used to predict the intermittent stall dynamics for different angles of attack. 
As shown in Fig.~\ref{fig:residence_time}, the evolution with $\alpha$  of the mean lifetimes $t_{res}$ of states A and D obtained by long temporal simulations of the identified model is in good agreement with  experimental measurements over two decades
(see Supplemental Material [\textit{URL will be inserted by publisher}] for details about the calculation of $t_{res}$).
We note that standard estimations of $t_{res}$ from the Arrhenius law and Eyring-Kramers formula cannot be used here because the noise intensity is not small compared to the potential barrier  height \cite{risken_fokkerplanck_1984}.

\textit{Discussion: saddle-node bifurcations} -- The identified model predicts the existence of an unstable deterministic branch, corresponding to the local maximum of $\phi$ in Figs.~\ref{fig:pdf}$(c)$ and \ref{fig:s-shape}$(d)$ and, physically, to flow state(s) visited transiently during the random switches between the stable states A and D.
In addition, a series expansion of the model around the turning points of the S-curve makes it possible to predict two local saddle-node bifurcations in $\alpha=\alpha_{A,D}$. 
More precisely,  for an equilibrium $F=0$ in $(C_L,\alpha)=(C_{L0},\alpha_0)$ 
where $\partial_{C_L} F=0$, $\partial_{C_L C_L} F \neq 0$ and $\partial_{\alpha} F \neq 0$, the system is locally topologically equivalent to a saddle-node bifurcation of normal form
\begin{equation}
\dot \eta = \beta + s \eta^2
\label{eq:nf}
\end{equation}
for the transformed variables 
\begin{align}
\eta = m_1 \left( \widetilde C_L + m_2 \widetilde \alpha \right),
\quad
\beta = m_1 m_3 \widetilde \alpha, 
\end{align}
where
$s = \mbox{sign}( \partial_{C_L C_L} F )$, 
$\widetilde C_L = C_L-C_{L0}$,
$\widetilde \alpha = \alpha-\alpha_{0}$,
$m_1 = \left| \partial_{C_L C_L} F \right|/2$, 
$m_2 = \partial_{\alpha C_L } F / \partial_{C_L C_L} F$, 
and
$m_3 = \partial_\alpha F$ \cite{Kuznetsov}.
Equilibrium solutions of (\ref{eq:nf}) are
\begin{equation}
\eta=\pm\sqrt{ -s \beta}
\label{eq:sd}
\end{equation}
for $s\beta<0$, and the stable/unstable branches are $s\eta<0$ and $s\eta>0$, respectively, as shown in Fig.~\ref{fig:s-shape}$(b)$-$(c)$.
The physical consequence of this result is that the D state cannot exist for angles smaller than $\alpha_D$, while the A state ceases to exist beyond $\alpha_A$, rather than simply becoming unstable. These theoretical predictions would benefit from targeted experimental campaigns to put them to the test. 
A qualitatively similar bifurcation scenario was observed in \cite{Busquet_2021} for a thin cambered airfoil at $Re=5\times 10^5$, where a deterministic S-shaped $C_L(\alpha)$ was obtained with a continuation method as the steady solution of a URANS model.

\textit{Discussion: physical meaning of the effective noise intensity} -- One may wonder what physically sets the effective noise intensity in the reduced-order model. 
While it seems sensible that the freestream turbulence intensity should have some influence on $\Gamma$, the exact mechanism remains unclear, and the flow state itself may play a role too.
Some insight can be borrowed from the laminar regime:
in the recent study \cite{Ducimetiere2024}, a stochastic amplitude equation similar to (\ref{eq:SDE_CL}) was  derived rigorously for a laminar flow undergoing a deterministic pitchfork bifurcation; in the presence of a stochastic forcing of scalar amplitude $F$ and spatial structure $\mathbf{f}(\mathbf{x})$, the effective noise intensity can be computed exactly and is found to be proportional to $F 
\langle \mathbf{u}^\dagger | \mathbf{f} \rangle / \langle \mathbf{u}^\dagger | \mathbf{u} \rangle,$
where $\mathbf{u}(\mathbf{x})$ and $\mathbf{u}^\dagger(\mathbf{x})$ are the  bifurcating eigenmode and corresponding adjoint mode. 
In other words, the effective noise intensity $\Gamma$ in the reduced-order model depends not only on the the stochastic forcing's amplitude $F$, but also on how well its spatial structure $\mathbf{f}$ aligns with the unstable mode.
Extending this result to the turbulent regime is a challenging task which, to this day, remains hypothetical.
Yet, some key aspects may be expected to qualitatively carry over, namely that the spatial projection between the forcing $\mathbf{f}$ and some mode $\mathbf{u}$ is crucial. 
If that mode, yet to be identified, is related to the mean flow, this could explain why we found  very different effective noise intensities $\Gamma_A$ and $\Gamma_D$: as the two mean states A and D differ widely, their respective modes $\mathbf{u}$ may differ too and so would their alignment with $\mathbf{f}$. 

We hypothesize that increasing the incoming turbulence intensity would not only lead to a larger $F$, but also modify the mean states and therefore the respective modes $\uu$ and the inner product $\langle \mathbf{u}^\dagger | \mathbf{f} \rangle$.
Although the theoretical framework has not yet been formalized in the turbulent regime, it would be insightful to conduct dedicated experimental or numerical studies aiming at characterizing the unstable eigenmodes of states A and D for different angles of attack and turbulence intensities.

\textit{Conclusion} -- 
In this Letter, we have derived a Langevin equation describing the bistable dynamics of an airfoil under varying angle of attack.
We obtained a good agreement both for the system's statistics (probability density functions) and for its dynamics (residence time). 
Key to this successful identification was a multiplicative (state-dependent) stochastic forcing. 
From the reduced-order model, we found the bistable region to be delimited by two saddle-node bifurcations.

We expect the proposed identification method to be useful for other aerodynamic flows exhibiting bistability between two widely different states (e.g. attached and separated), but also for a variety of other parameterized, multistable dynamical systems   
subject to multiplicative stochastic forcing, thereby facilitating model reduction and flow control in many physical fields. 

Whether Langevin equations can be derived from first principles, not only  for stochastically forced laminar flows \cite{Ducimetiere2024} but also for intermittent turbulent flows, remains an open question.
Promising  directions for the study of intermittent dynamics include computational approaches based on unstable solutions (e.g. equilibria, periodic orbits, tori) \cite{KAWAHARA_KIDA_2001, Chandler_Kerswell_2013, Budanur_etal_2017, Suri2024} 
and most probable transition paths \cite{Lecoanet2018, PhysRevLett.122.074502, Gome2022}.

\bigskip
\textit{Acknowledgments} -- 
The authors thank the anonymous referees for insightful questions and suggestions, as well as Lebo Molefe for a critical reading of  the manuscript.
E.~B. is grateful to Yves-Marie Ducimetière and Fran\c{c}ois Gallaire for interesting discussions.
 
\bibliography{ALL2}

\clearpage
\newpage

\onecolumngrid

\begin{center}
\large{
\textbf{Bistable flow dynamics of airfoil stall under varying angle of attack:
}}

\large{
\textbf{A stochastic model with multiplicative noise\\
\textit{Supplemental Material}}
}

\bigskip 

\normalsize
Edouard Boujo

\textit{\small{Laboratory of Fluid Mechanics and Instabilities, École Polytechnique Fédérale de Lausanne, CH-1015 Lausanne, Switzerland} 
}

\bigskip 

\normalsize
Ivan Kharsansky Atallah

\textit{\small{Fluid Mechanics Department, ENSTA Paris, Institut Polytechnique de Paris, F-91120 Palaiseau, France and} }

\textit{\small{EM2C Laboratory, CNRS, CentraleSupelec, Universit\'e Paris-Saclay, F-91190 Gif-sur-Yvette, France} }

\bigskip 

\normalsize
Luc R. Pastur

\textit{\small{Fluid Mechanics Department, ENSTA Paris, Institut Polytechnique de Paris, F-91120 Palaiseau, France} }
\end{center}

\normalsize

\bigskip
\bigskip
\bigskip

In this document, we provide additional details about the Markov property of our experimental lift measurements, the need for multiplicative noise in our reduced-order model, and the calculation of the mean residence times from the experimental data and from the identified reduced-order model.

\bigskip
\bigskip
\bigskip

\begin{center}
\textbf{Markov property}
\end{center}

\begin{figure}
\centerline{  
  \hspace{0.72cm}
  \begin{overpic}[height=5.5cm, trim=0mm 0mm 0mm 8mm, clip=false]{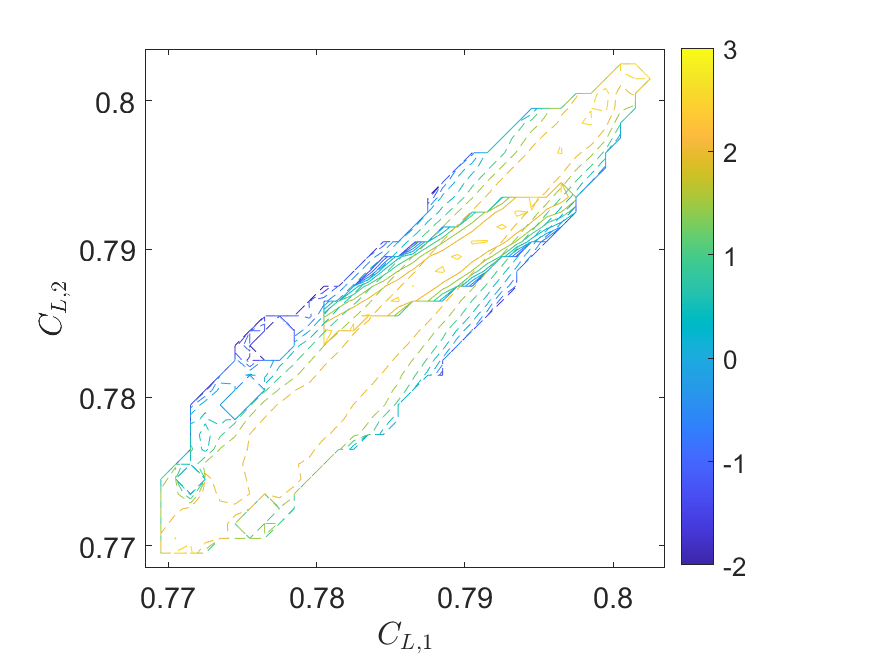}
     \put(4,69){$(a)$}
  \end{overpic}  
  \begin{overpic}[height=5.5cm, trim=0mm 0mm 0mm 8mm, clip=false]{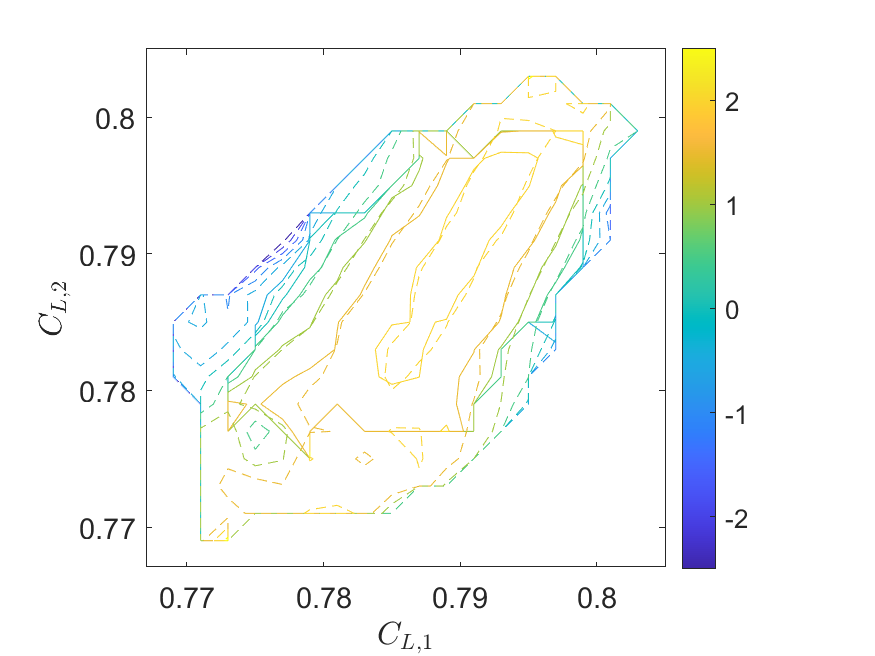}
     \put(4,69){$(b)$}
  \end{overpic}  
}
\vspace{0.4cm}
\centerline{  
  \begin{overpic}[height=5.5cm, trim=0mm 0mm 0mm 8mm, clip=false]{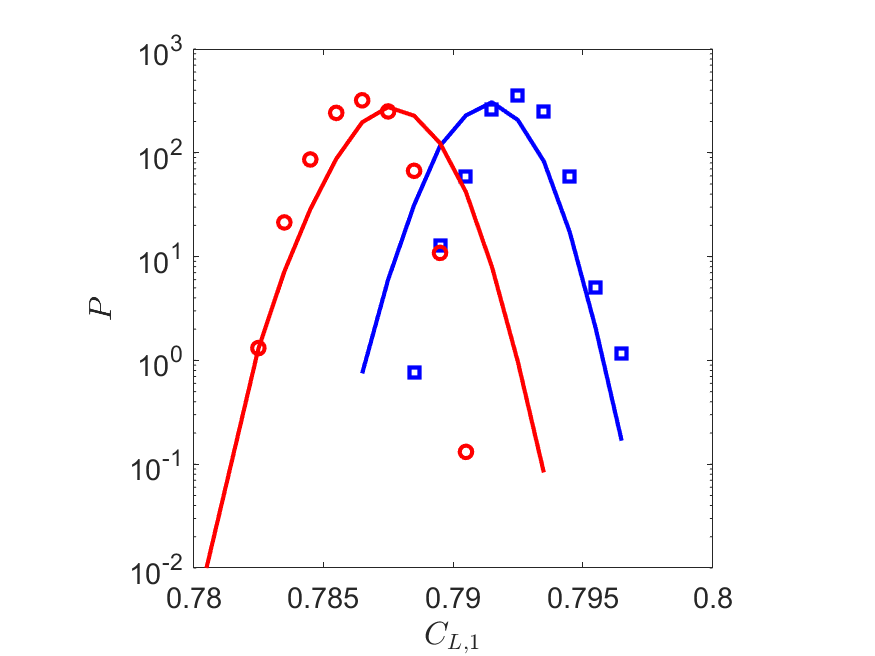}
  \end{overpic}  
  \begin{overpic}[height=5.5cm, trim=0mm 0mm 0mm 8mm, clip=false]{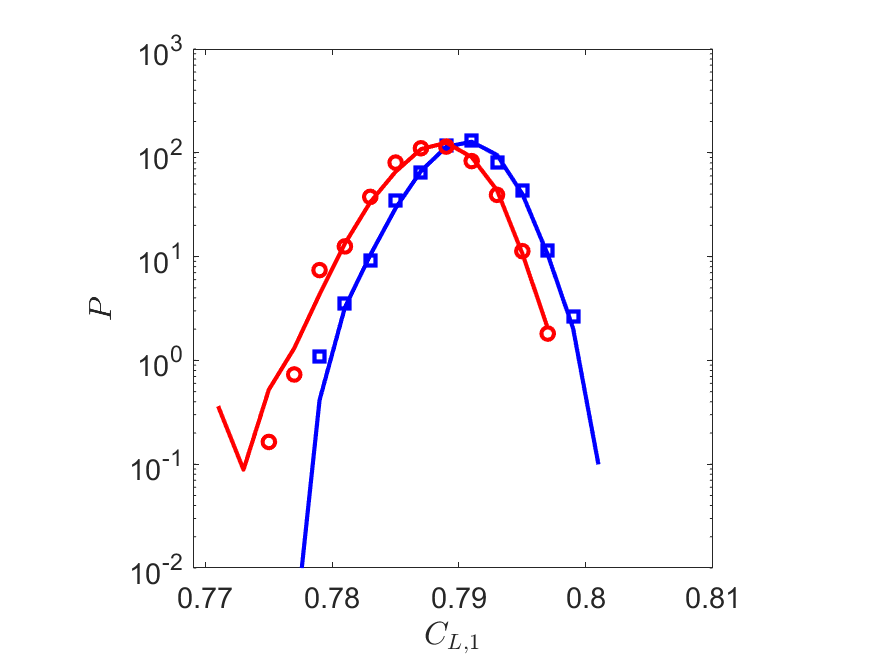}
  \end{overpic}  
}
\caption{
Conditional PDFs $P(C_{L,1},t_1 | C_{L,2},t_2)$ and $P(C_{L,1},t_1 | C_{L,2},t_2; C_{L,3},t_3)$ for $t_3-t_2=t_2-t_1=\Delta t $, with
$(a)$~$\Delta t=0.3$~s and 
$(b)$~$\Delta t=1.0$~s.
Contours (logarithmic scale) show $P(C_{L,1},t_1 | C_{L,2},t_2)$ as dashed lines and $P(C_{L,1},t_1 | C_{L,2},t_2; C_{L,3},t_3)$ as solid lines;
cuts in $C_{L,2}=0.7877$ (red) and $C_{L,2}=0.7913$ (blue) show $P(C_{L,1},t_1 | C_{L,2},t_2)$ as solid lines and $P(C_{L,1},t_1 | C_{L,2},t_2; C_{L,3},t_3)$ as symbols.
$\alpha= 10.70^\circ$. 
$C_{L,3} = 0.789$. 
}
\label{fig:Markov}
\end{figure}

To demonstrate that the stochastic process is a Markov process, we perform a direct test of the Markov property 
$$ P(C_{L,1},t_1 | C_{L,2},t_2; C_{L,3},t_3; \ldots; C_{L,N},t_N)= P(C_{L,1},t_1 | C_{L,2},t_2) $$
for $N=3$ \cite{FRIEDRICH201187}: we compute the one-time  conditional probability 
$$P(C_{L,1},t_1 | C_{L,2},t_2)$$ 
and the two-time conditional probability 
$$P(C_{L,1},t_1 | C_{L,2},t_2; C_{L,3},t_3).$$
Fig.~\ref{fig:Markov} shows contour plots of these conditional probabilities, together with cuts  at fixed values of $C_{L,2}$, for the angle of attack $\alpha= 10.70^\circ$, representative of the high-lift regime. 
(We obtain similar results for $\alpha=11.02^\circ$, representative of the low-lift regime.)
We set the value of $C_{L,3}$ close to the most probable lift coefficient, which here is 0.789. 
We also test several values of the time difference $\Delta t= t_3-t_2=t_2-t_1$. 
The proximity of the contours and cuts in panel $(b)$ ($\Delta t=1.0$~s) shows that the Markov property holds if $\Delta t$ is large enough, while panel $(a)$ ($\Delta t=0.3$~s) shows that this property is lost if $\Delta t$ is too small. 
Similar conclusions are drawn from cuts in $C_{L,2}=0.7877$ (red) and $C_{L,2}=0.7913$ (blue).
This is consistent with the signals being low-pass filtered at $f_{LP} = 1$~Hz \cite{kharsansky2024}, implying that $\Delta t$ should not be chosen much smaller  than $1/f_{LP}=1$~s.
In the identification, we only used time shifts $\tau \geq 1$~s, a scale at which the Markov property holds.

\clearpage
\newpage

\begin{center}
\textbf{Need for multiplicative noise}
\end{center}

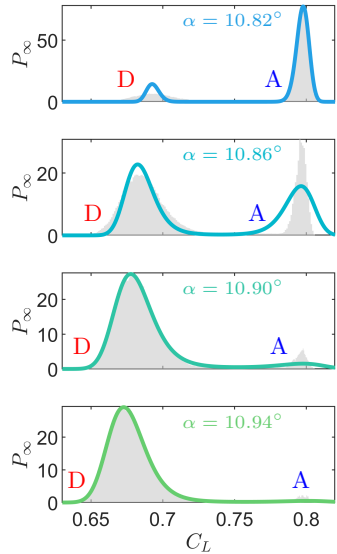
\begin{figure}[h]
\vspace{0.4cm}
\centerline{
  \begin{overpic}[height=7.273cm, trim=0mm 0mm 0mm 0mm, clip=true]{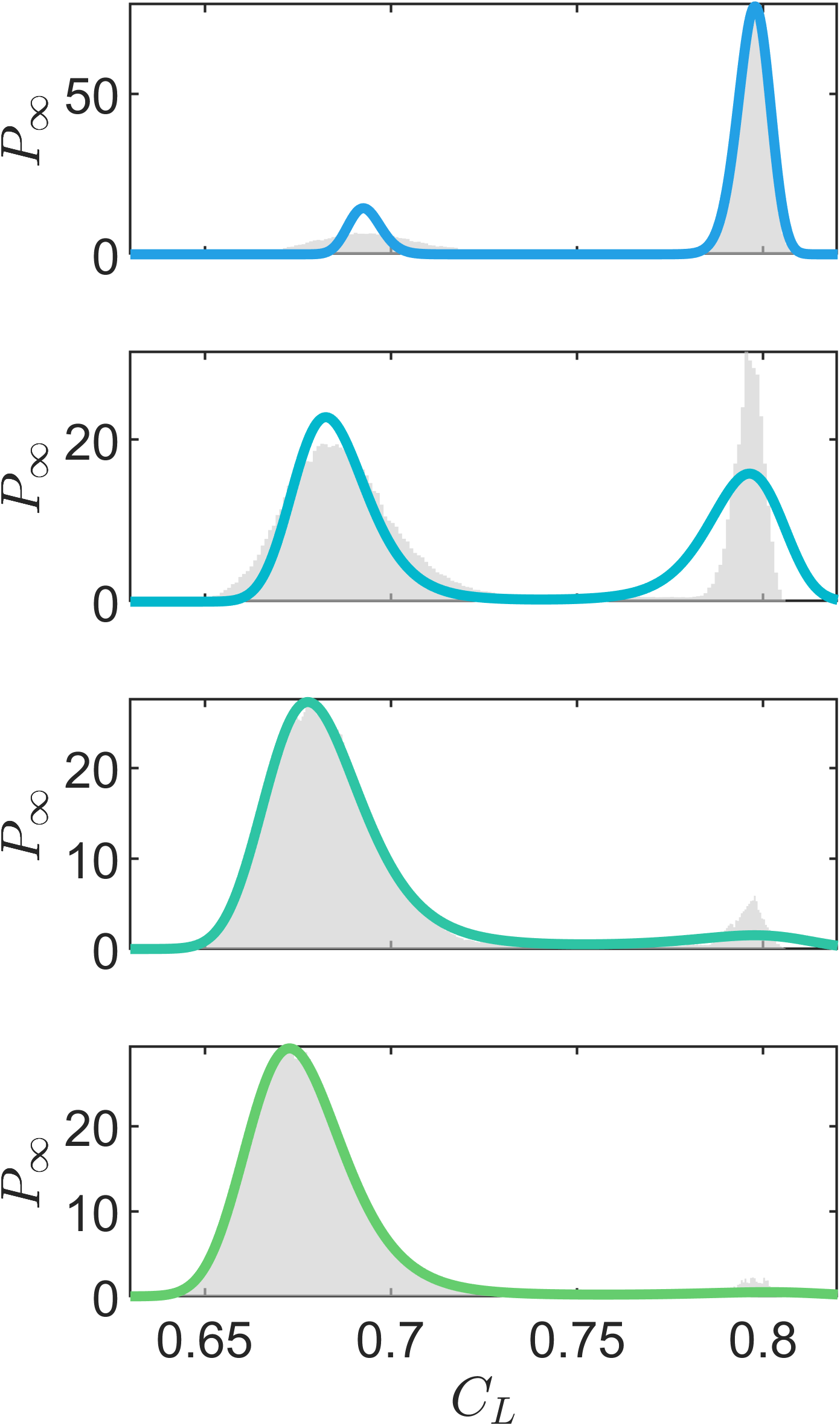}
     \put(31,96  ){\textcolor[rgb]{0.1390   0.6259   0.8981}{\scriptsize$\alpha=10.82^\circ$}}
     \put(31,71.5){\textcolor[rgb]{0.0097   0.7160   0.8009}{\scriptsize$\alpha=10.86^\circ$}}
     \put(31,47  ){\textcolor[rgb]{0.1804   0.7701   0.6447}{\scriptsize$\alpha=10.90^\circ$}}
     \put(31,22.5){\textcolor[rgb]{0.3956   0.8029   0.4314}{\scriptsize$\alpha=10.94^\circ$}}
     \put(19,  85)  {\textcolor{red}{D}}
     \put(13,  60.5){\textcolor{red}{D}}
     \put(11,  36)  {\textcolor{red}{D}}
     \put(10,  11.5){\textcolor{red}{D}}
     \put(46, 85)  {\textcolor{blue}{A}}
     \put(43, 60.5){\textcolor{blue}{A}}
     \put(47, 36)  {\textcolor{blue}{A}}
     \put(51, 11.5){\textcolor{blue}{A}}
  \end{overpic} 
}
\caption{
Lift PDF for a few angles of attack in the bistable region. 
Shaded histogram: experimental $P_\infty(C_L)$.
Solid line: individual fit $P^a_\infty(C_L)$ with additive noise (constant $\Gamma$).
}
\label{fig:pdf_additive} 
\end{figure}

Figure \ref{fig:pdf_additive}  shows the best fit of the lift PDF obtained when assuming additive noise, i.e. using Eq.~(7) of the main text instead of Eq.~(5). 
Clearly, and unlike Fig.~3 of the main text, the agreement is poor in the bistable region: the additive-noise model cannot simultaneously capture the large fluctuations of state D and smaller fluctuations of state A, thus motivating the use of multiplicative noise.

\bigskip

\begin{center}
\textbf{Mean residence times}
\end{center}

Mean residence times $t_{res}$ are computed as follows.
On the experimental side, measured time signals $C_L(t)$ are processed to detect transition events between states A and D, defined as $C_L(t)$ crossing the local minimum of $P_\infty$ (local maximum of $\phi$). 
The residence time $t_{res}^{\mathrm{A}}$ is the time spent in state A, i.e. between two successive transitions  D$\rightarrow$A and A$\rightarrow$D, and vice-versa for $t_{res}^{\mathrm{D}}$. 
Averaging all the residence times $t_{res}^{\mathrm{A}}$ and $t_{res}^{\mathrm{D}}$ yields the respective mean residence times.
On the model side, numerical time signals $C_L(t)$ are generated by solving in time the Langevin equation (1) of the main text, with the drift coefficient $F(C_L)$ deriving from the potential $V(C_L)$ given by (2), the identified coefficients $a(\alpha)$, $b(\alpha)$, $c(\alpha)$ and $d(\alpha)$ given by (10), and the noise $\Gamma(C_L)$ given by (3).
We use the Euler-Maruyama method with a time step of $10^{-3}$~s and a total time $T=5\times10^4$~s. The same processing as for the experimental signals is then applied to obtain the mean residence times $\langle t_{res}^{\mathrm{A}}\rangle$ and $\langle t_{res}^{\mathrm{D}}\rangle$.

\end{document}